\documentclass[letterpaper,10pt,conference]{ieeeconf}  

\IEEEoverridecommandlockouts                              

\usepackage{graphics} 
\usepackage{epsfig} 
\usepackage{amsmath} 
\usepackage{amssymb}  
 
\usepackage{amsthm}
\usepackage{cite}
\usepackage{color}
\usepackage{enumerate}
\usepackage{graphicx}
\usepackage{mathtools}
\usepackage{cancel}
\usepackage{url}

\usepackage{subfigure}
\usepackage{graphicx}

\DeclareMathOperator*{\argmin}{argmin}

\newcommand{\naturals}{\mathbb{N}}
\newcommand{\real}{\mathbb{R}}
\newcommand{\realnonneg}{\mathbb{R}_{\ge 0}}
\newcommand{\realpos}{\mathbb{R}_{> 0}}

\newcommand{\map}[3]{#1:#2 \rightarrow #3}

\newcommand{\longthmtitle}[1]{\mbox{}{\textit{(#1):}}}
\newcommand{\setdef}[2]{\{#1 \; | \; #2\}}

\newcommand*{\SetSuchThat}[1][]{} 
\newcommand*{\MvertSets}{%
    \renewcommand*\SetSuchThat[1][]{%
        \mathclose{}%
        \nonscript\;##1\vert\penalty\relpenalty\nonscript\;%
        \mathopen{}%
    }%
}
\MvertSets 
\DeclarePairedDelimiterX \Set [2] {\lbrace}{\rbrace}
    {\,#1\SetSuchThat[\delimsize]#2\,}

\newcommand{\Cc}{\mathcal{C}}
\newcommand{\Dc}{\mathcal{D}}

\newcommand{\Kc}{\mathcal{K}}

\newcommand{\Sc}{\mathcal{S}}

\newcommand{\emm}{e_{\rm{max}}}
\newcommand{\ie}{i.e., }
\newcommand{\defeq}{\triangleq}
\newcommand{\de}{\mathcal{\delta}}

\newtheorem{theorem}{Theorem}
\newtheorem{lemma}{Lemma}

\theoremstyle{definition}
\newtheorem{definition}{Definition}

\newtheorem{remark}{Remark}

\newtheorem{assumption}{Assumption}

\newcommand{\vr}{\varepsilon}
\newcommand{\nom}{{\operatorname{nom}}}
\newcommand{\m}{{\operatorname{min}}}

\newcommand{\Lie}{\mathcal{L}}

\renewcommand{\baselinestretch}{1}
\allowdisplaybreaks
\title{\LARGE \bf
Sample-and-Hold Safety with Control Barrier Functions 
}

\author{Gilbert Bahati, Pio Ong, and Aaron D. Ames
\thanks{This research was supported in part by the National Science Foundation (CPS Award \#1932091) and the Technology Innovation Institute (TII).}
\thanks{The authors are with the Department of Mechanical and Civil Engineering, California Institute of Technology, Pasadena, CA 91125, USA. {\tt\small \{gbahati,pioong,ames\}@caltech.edu}}%
}

\begin{document}

\maketitle
\thispagestyle{empty}
\pagestyle{empty}

\begin{abstract}
A common assumption on the deployment of safeguarding controllers on the digital platform is that high sampling frequency translates to a small violation of safety. This paper investigates and formalizes this assumption through the lens of Input-to-State Safety. From this perspective, and leveraging control barrier functions (CBFs), we propose an alternative solution for maintaining safety of sample-and-hold controlled systems without any violation to the original safe set. Our approach centers around modulating the sampled control input in order to guarantee a more robust safety condition.  We analyze both the time-triggered and the event-triggered sample-and-hold implementations, including the characterization of sampling frequency requirements and trigger conditions. We demonstrate the effectiveness of our approach in the context of adaptive cruise control through simulations.
\end{abstract}


\section{Introduction}
Safety-critical systems are a crucial part of a variety of application domains: transportation, manufacturing, healthcare among others. In such systems, the occurrence of safety violations, no matter how small, may lead to catastrophic consequences. In these systems, electronic devices are often used to implement digital control, where continuous signals from the real-world systems are sampled at regular intervals \cite{MJMP:10}. However, this can result in measurement uncertainties due to sampling (i.e., zero-order hold errors), which can lead to safety violations. Therefore, it is essential to design robust control methods that can account for these errors in order to guarantee safety.
To mitigate these sample-and-hold errors, a common practice is to sample at high frequencies, limiting the error to a small and hopefully negligible value. However, this does not entirely eliminate the error, only reducing it to an acceptable level. It is often still the case that absolute safety (even in the presence of these errors) is desired. 

To avoid dealing with the sample-and-hold error explicitly, it is common to consider a discrete time approximation~\cite{AA-SK:17} of the available continuous time system as done in sampled data systems~\cite{AT-VD-RK-YY-AA:22, shaw_cortez_control_2021,AG-IA-DR-SL-MK:18, JB-KG-PD:21,LN-ZH-AC:21,YZ-SW-XX:22 }. However, obtaining these approximations can introduce additional error since the exact solutions to the continuous time systems are generally not known. Furthermore, discrete time designs explicitly guarantee safety only at sampling instants (when the controller is updated) and not in the inter-sample periods (when the control signal is being held constant) and as such, safety may be violated during this period. To formalize this property of discrete time approximations, the notion of practical safety~\cite{AT-VD-RK-YY-AA:22} is usually adopted to account for these possible violations during the inter-sample period. Practically safe systems can maintain their state within a safe set during sample times, provided that the sampling frequency is high enough. This results in safety being ensured with respect to a larger set that accommodates all possible violations of the original set.

The notion of safe set expansion to accommodate uncertainties in the system can be encapsulated by Input-to-State Safety (ISSf)\cite{SK-ADA:18}.  In particular, this generalizes Input-to-State Stability (ISS)~\cite{sontag_input--state_1995,jiang_input--state_2001,freeman2008robust} in the context of safe sets, wherein disturbances in the input result in convergence to a region of the safe set dictated by the size of the disturbance. 
To certify ISSf, the framework of Control Barrier Functions (CBFs) \cite{ADA-XX-JWG-PT:17,ADA-SC-ME-GN-KS-PT:19,jankovic2018robust} can be leveraged.  In particular, the existence of an ISSf-CBF implies the system is ISSf.  This notion was generalized in a ``tunable'' fashion in \cite{AA-AJT-CRH-GO-AA:22} allowing for more control over the expansion of the safe set---an extension that has found applications in practice \cite{cosner2022safety,alan2022control}.  
Yet obtaining safety guarantees in the context of CBFs does not address digital implementation issues raised by sample-and-hold.

There is a rich body of work on translating discretely instantiated continuous-time systems in a way that preserves desired properties, encapsulated by the area of event-triggered control \cite{PT:07,WPMHH-KHJ-PT:12}.  This has historically studied stability, and used Lyapunov functions to quantify when to sample and hold so as to guarantee that stability is maintained: the result are Lyapunov-based trigger laws that utilize ISS to bound the system behavior during the inter-sample periods \cite{PT:07}.  Event-triggered control has also been studied in the context of safety using CBFs \cite{AJT-PO-JC-AA:21-csl,long2022safety,KS-DC-MK:22,XW-BC-CC:22}. But these instantiations required expansion of the safe set to accommodate inter-sample errors.   Therefore, the end result are small safety violations much like what is seen in the sample-and-hold framework. 

The main contribution of this paper is a robust control framework for sample-and-hold control systems that establishes safety without violations or relaxations of the safe set of interest. We achieve this by mathematically formalizing the notion of practical safety \cite{AT-VD-RK-YY-AA:22} through the Input-to-State Safety (ISSf) framework \cite{SK-ADA:18}.
In particular, we present a modulation of the nominal controller that guarantees safety on the original safe set without resorting to expansions of the set. Similar nominal controller adjustments have been proposed in the literature to provide ISSf guarantees to a system  \cite{SK-ADA:18,AA-AJT-CRH-GO-AA:22}; however, to the best of our knowledge, these approaches have not been utilized in the context of sample-and-hold, wherein new complications arise. Furthermore, to mitigate the conservative nature of the proposed designs, we leverage the event-triggered control approach \cite{PT:07, AJT-PO-JC-AA:21-csl}  to synthesize controllers that address such conservatism yet still maintain robustness to sample-and-hold errors.  In short, the main result guarantees the safety (forward invariance) of the original safe set---attenuating conservatism while ensuring  inter-sample safety.

\noindent \subsubsection*{Notation}
We utilize $\naturals$, $\real$, $\realnonneg$, $\realpos$ to denote natural, real, nonnegative, and positive numbers, respectively. For a vector $x\in\real^n$, $\|x\|$ denotes the Euclidean norm and $\|x\|_\Cc = \inf\setdef{\|x-y\|}{y\in\Cc}$ is its Hausdorff distance to set $\Cc$. Then given a set $\Cc$, a $\delta$-neighborhood of $\Cc$ is the set $\setdef{x}{\|x\|_\Cc < \delta}$. A function $\map{\alpha}{[0,a)}{\realnonneg}$ with $a>0$, is of class-$\Kc_\infty$ if $\alpha(0)=0$, $\alpha$ is strictly increasing, and $\lim_{s\rightarrow\infty} \alpha(s) = \infty$. A function $\map{\beta}{[-b,a)}{\real}$ is of class-$\Kc_\infty^e$ if it is of class-$\Kc_\infty$ and $\beta(y) \leq 0$ for all $y \in [-b,0]$.



\section{Safety For Continuous-Time Systems} \label{Background Safety}
This section provides the background on barrier functions and its usage for guaranteeing safety for continuous-time systems. In addition, we will review the robustness concept in safety within the construct of Input-to-State Safety. 

\subsubsection*{Safety For Continuous Time Systems}
We consider the nonlinear system:
\begin{equation}\label{eq:NL}
\dot x = F(x,u),
\end{equation}
where $x \in \mathbb{R}^n$ is the state and $u \in \mathbb{R}^m$ is the input. The system vector field $\map{F}{\real^n\times \real^m}{\real^n}$ is assumed to be locally Lipschitz. 
The goal of safety is to contain all system trajectories within a \textit{safe set} $\Cc\subset \real^n$. In other words, we want the set $\Cc$ to be forward invariant.

\begin{definition}\longthmtitle{Safety} A set $\Cc$ is forward invariant if for every initial condition $x_0 \in \Cc$, its ensuing trajectory $x(t) \in \Cc$ $\forall \, t \geq 0$. A system is safe on $\Cc$ if the set $\Cc$ is forward invariant. \hfill$\bullet$
\end{definition}
One approach to guaranteeing safety is to describe a given set $\Cc$ with a continuously differentiable \textit{barrier function} $h : \mathbb{R}^n \rightarrow \mathbb{R}$ such that:
\begin{subequations}\label{eq:barrier}
\begin{align}
\Cc = \{x \in \mathbb{R}^n : h(x) \geq 0\}, \\
\partial \Cc = \{x \in \mathbb{R}^n : h(x) = 0\}, \\
\text{Int}(\Cc) = \{x \in \mathbb{R}^n : h(x) > 0\}.
\end{align}
\end{subequations} 
In this case, the set $\Cc$ is the 0-superlevel set of the barrier function~$h$, and the safety goal is then to keep the function $h$ positive at all times. This can be achieved by designing a Lipschitz continuous controller $\map{k}{\real^n}{\real^m}$ so that the state feedback $u=k(x)$ satisfies the \textit{barrier condition} \cite{ADA-XX-JWG-PT:17}:
\begin{equation}\label{eq:CBF}
 \frac{\partial h}{\partial x}(x)F(x,k(x)) \geq -\alpha(h(x)),
\end{equation}
for some class-$\Kc_\infty^e$ function $\alpha$. When the above condition is met, we call the controller $k$ a \textit{safeguarding controller} that renders the system safe on $\Cc$.

The barrier condition provides a conservative approach for achieving safety. In addition to making sure that on the boundary of the safe set, the vector field of the closed-loop system points towards the interior of the set (cf., Nagumo's Theorem~\cite{FB:99}), the condition also limits the speed at which the trajectories may approach the boundary~$\partial\Cc$. One benefit of this approach is the ability to formulate robustness through the notion of Input-to-State Safety, which we discuss next.

\subsubsection*{Safety Robustness}
In the presence of disturbances added to the system, a safeguarding controller may fail to render the system safe on the safe set $\Cc$. In such cases, we may want the system trajectories to at least stay close to the safe set.
One mathematical framework for guaranteeing such a behavior is Input-to-State Safety\footnote{Sometimes more precisely as Disturbance-Input-to-State Safety}.

Consider the closed-loop nonlinear control-affine system with disturbance:
\begin{equation}\label{eq:NL distrubed}
\dot{x} = F_e(x,k(x), e),
\end{equation}
where $e\in\real^{n_e}$ is a bounded disturbance. We assume $\map{F_e}{\real^n\times\real^m\times\real^{n_e}}{\real^n}$ is locally Lipschitz in the state and control, and continuous in disturbance. Input-to-State Safety seeks to establish a relationship between the maximum size that the disturbance signal $t\mapsto e(t)$ achieved along the trajectory (i.e., ~$\| e\|_\infty$), and how far the trajectory may stray away from the safe set. Given a safe set $\Cc$, we use the barrier function $h$ as a proxy to define a larger set $\Cc_{e} \supset \Cc$ as:
\begin{subequations}\label{eq:barrierISS}
\begin{align}
\Cc_{e} = \{x \in \mathbb{R}^n : h(x) + \gamma(\|e\|_{\infty})\geq 0\}, \\
\partial \Cc_{e} = \{x \in \mathbb{R}^n : h(x) + \gamma(\|e\|_{\infty}) = 0\}, \\
\text{Int}(\Cc_{e}) = \{x \in \mathbb{R}^n : h(x) + \gamma(\|e\|_{\infty}) > 0\},
\end{align}
\end{subequations}
where $\gamma$ is a class-$\Kc_\infty$ function. 
\begin{definition}\longthmtitle{Input-to-State Safety}
    A system is Input-to-State Safe (ISSf) on  $\Cc$ if there exists a function $\gamma \in \Kc_\infty$ such that it is safe on $\Cc_{e}$ under any disturbance signal $t\mapsto e(t)$.~\hfill$\bullet$
\end{definition}
The existence of a function $\gamma$, and therefore Input-to-State Safety on set $\Cc$, can be established if the following \textit{Input-to-State Safe barrier condition} (ISSf-BC)~\cite{SK-ADA:18} holds:
\begin{equation}\label{eq:ISSf-CBF}
\frac{\partial h}{\partial x}(x)F_e(x,k(x),e)   \geq -\alpha(h(x)) - \iota(\|e\|)
\end{equation}
for some class-$\Kc_\infty$ function $\iota$.

\section{Practical Safety from ISSf Perspective}
We focus on discretely instantiated continuous-time systems that result from sample-and-hold (i.e., zero-order hold) implementations of safeguarding controllers on digital platforms, and examine potential safety issues that arise. We limit ourselves to control-affine systems:
\begin{equation}\label{eq:ctrl-affine}
    \dot x = f(x)+g(x)u,
\end{equation}
where the system vector fields $f:\mathbb{R}^n \rightarrow \mathbb{R}^n$ and $g:\mathbb{R}^n \rightarrow \mathbb{R}^{n \times m}$ are assumed to be locally Lipschitz. 

Given a nominal controller $\map{k_\nom}{\real^n}{\real^m}$, the sample-and-hold implementation strategy is as follows: the controller is sampled at a time instant $t_i$, then the control value $u=k_\nom(x(t_i))$ is held constant until $t_{i+1}$, the next sampling instant. As a result, the closed-loop system is given by:
\begin{equation}\label{eq:NL_sampled}
\dot x = f(x) + g(x)k_\nom(x+e) \ \  \forall \, t \in [t_i, t_{i+1}),
 \end{equation}  
with a sample-and-hold error $e = x(t_i)-x$. Due to the presence of this sample-and-hold error, the safeguarding nominal controller may not be able to fulfill its safety task, i.e., rendering the sample-and-hold system~\eqref{eq:NL_sampled} safe on~$\Cc$.

\subsection{Practical Safety}\label{Set Expansion Section}

In practice, a typical approach for dealing with sample-and-hold errors is through high frequency sampling, relying on the idea referred to as \textit{practical safety}~\cite{AT-VD-RK-YY-AA:22}.

\begin{figure}[t!]
\centering
\includegraphics[width=1\linewidth,height=\textheight,keepaspectratio]{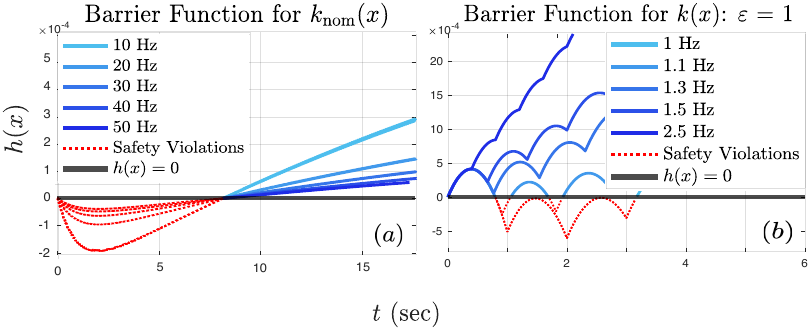}
\caption{(a) Safety under the nominal controller via sample-and-hold.  Here, we see violations regardless of the sampling frequency (b) Robustness of the controller adjustment presented in Section \ref{violation free section}. Here, we satisfy safety even for slow sampling frequencies. Both (a) and (b) are initialized from $h(x) = 0$ (\ie the set boundary) and correspond to the adaptive cruise control system presented in Section \ref{Application}.}
\label{fig practical safety}
\vspace{-5mm}
\end{figure}

A system is deemed practically safe if the states (along the trajectories starting from $\Cc$), can be maintained at a distance arbitrarily close to the original safe set $\Cc$, with a sufficiently high sampling frequency, i.e., violations of safety appear to be minor if we sample fast enough, as illustrated in Figure~\ref{fig practical safety}.

Although the description of practical safety seems reasonable, it is quite difficult to show that a controller yields practical safety. 

To this end, the usual argument for practical safety is that of Input-to-State safety (ISSf) discussed in Section~\ref{Background Safety}. The idea is based on two key reasonable assumptions:

\begin{itemize}
    \item \textbf{Minor Safety Violations:} Small errors~$e$ should only lead to minor violations of safety of $\Cc$. 
    \item \textbf{Small Error Bounds:} Under high frequency sampling, the error~$e$ should not be able to grow too large. 
\end{itemize}

Although these concepts are often cited as the reasons for practical safety, to the best of our knowledge, no mathematical formalization of them exists. In the subsequent section, we develop a set of assumptions that enable us to use ISSf as a basis for practical safety.

\subsection{Minor Safety Violations}

Practical safety and Input-to-State Safety (ISSf) both acknowledge the fact that the presence of disturbances in a system may harm the safety guarantees provided by a controller. Furthermore, both aim to characterize safety violations as an increasing function of the size of disturbances. However, the two concepts are not exactly the same. ISSf describes safety violations with superlevel sets of barrier functions rather than generally with any set expansion. On the other hand, practical safety particularly deals with set expansions arising from the effect of sample-and-hold errors. Nevertheless, there is a great overlap between the two concepts, and our work lies at their intersection. Specifically, we consider when the nominal controller is a safeguarding controller satisfying the barrier condition~\eqref{eq:barrier} for the nominal system~\eqref{eq:ctrl-affine} as:
\begin{align}\label{eq:CBF_ctrl-affine}
 \underbrace{\frac{\partial h}{\partial x}(x)f(x)}_{\Lie_fh(x)} + \underbrace{\frac{\partial h}{\partial x}(x)g(x)}_{\Lie_gh(x)}k_\nom(x) \nonumber \geq -\alpha(h(x)).
\end{align}
However, the evolution of the barrier function along the trajectory of the sample-and-hold system~\eqref{eq:NL_sampled} is given by:
\begin{equation}\label{eq:CBC_sampled}
 \dot{h}(x,k_\nom(x+e)) \defeq 
 \Lie_fh(x) + \Lie_gh(x) k_\nom(x+e),
 \end{equation}
 which may no longer satisfy the desirable barrier condition:
\begin{equation}\label{eq:unsafe sample-and-hold}
\dot{h}(x,k_\nom(x+e))\geq -\alpha(h(x)).
\end{equation}
Thus, safety of the sample-and-hold system on $\Cc$ cannot be guaranteed without an additional assumption accounting for sample-and-hold errors, which is why practical safety considers allowing minor safety violations. Similarly, ISSf establishes that small errors yield minor safety violations if there exists a function $\iota$ such that the ISSf-BC holds:
\begin{equation}\label{eq:CBF_ctrl-affine-sample-and-hold}
 \dot{h}(x,k_\nom(x+e)) \geq -\alpha(h(x)) -  \iota(\|e\|).
\end{equation}

The following regularity assumptions allow us to obtain~\eqref{eq:CBF_ctrl-affine-sample-and-hold}.

\begin{assumption}\label{Lipschitz controller}\longthmtitle{Lipschitz Dynamics} The functions $f, g$ and the controller $k$ are Lipschitz continuous on $\mathbb{R}^n$. ~\hfill$\bullet$
 \end{assumption}
\begin{assumption}\label{Lgh upper bound}
\longthmtitle{Upper Bound on $\Lie_gh$} The function $\Lie_gh$ is bounded above. That is, there exists a positive constant $\lambda > 0$ such that $\|\Lie_gh(x)\| \leq \lambda$ for all $x\in\Cc$. ~\hfill$\bullet$
 \end{assumption}
 \begin{remark}\longthmtitle{Regularity Assumptions on Region of Operation}
\label{rmk:regularity}
    Throughout this paper, we make regularity assumptions on bounds, as well as Lipschitzness, on different functions. We note that these terms are only required along the trajectories. Usually, there are multiple safety constraints (with possible additional stability constraints) in practice. Our bounds do not need to hold for all points in each safe set, but rather in their intersection where system trajectories are feasible, i.e., \textit{region of operation}. To this end, if the final \textit{region of operation} is compact, then all bounds exist, and Lipschitzness only needs to be local.\hfill $\bullet$
\end{remark}

Under Assumptions \ref{Lipschitz controller}-\ref{Lgh upper bound}, we derive the ISSf-BC \eqref{eq:CBF_ctrl-affine-sample-and-hold} as:
 \begin{align} \label{eq:ISSf-lambda-final}
 \dot{h}(x,k_\nom(x+e)) &= \Lie_fh(x) + \Lie_gh(x)k_\nom(x) \nonumber \\ 
    &\quad + \Lie_gh(x)\big(k_\nom(x(t_i)) - k_\nom(x)\big) \nonumber \\
    &\geq -\alpha(h(x)) - \underbrace{L_k\lambda \|e\|}_{\iota(\|e\|)},
\end{align}
for $x\in\Cc$, where $L_k >0$ is the Lipschitz constant of the controller. Typically, the ISSf-BC can be used to establish forward invariance of an expanded set $\Dc\supset\Cc$ in the form:
\begin{align}\label{eq:expanded set}
\Dc = \{x \in \mathbb{R}^n : h_d(x) = h(x) + d\geq 0\},
\end{align}
for some positive constant $d>0$ dictating the size of safety violation being accommodated. 
However, the bound $\lambda$ used in \eqref{eq:ISSf-lambda-final} only holds on the set $\Cc$ as stated in Assumption \ref{Lgh upper bound}. Therefore a new bound $\lambda' >\lambda$ must first be established on $\Dc$. Note importantly that we may not directly assume a bound $\lambda'$ on $\Dc$ because we use it to help define $\Dc$. To establish the existence of a $\lambda'$, we make another regularity assumption:
\begin{assumption}\longthmtitle{Lipschitz Continuity of $\Lie_gh$}\label{Lipschitz assumption Lgh-new} The function $\Lie_gh$ is Lipschitz continuous on $\real^n$. That is, $\exists \ M > 0$ such that $\|\Lie_gh(x)-\Lie_gh(y)\| \leq M\|x - y\| \ \forall \ x, y \in \Cc$.  ~\hfill$\bullet$
\end{assumption}
The above assumption allows us to characterize the behavior of $\Lie_gh$ locally in the neighborhood of $\Cc$. In particular, consider a $\de$-neighborhood of $\Cc$, denoted by $\Sc$. Because $\|x\|_\Cc<\delta$ for all $x\in\Sc$,
we can show that:
\begin{align}
    \|\Lie_gh(x)\| 
                &\leq \|\Lie_gh(x)- \Lie_gh(y)\|+\|\Lie_gh(y)\| \nonumber \\
                &\leq M\|x\|_\Cc+\lambda\nonumber \\
                &\leq M\de+\lambda = \lambda', \nonumber
\end{align}
for all $x\in\Sc$,
and the following property holds:
 \begin{align} \label{eq:lambda prime}
\dot{h}(x,k_\nom(x+e))\geq -\alpha(h(x)) - L_k\lambda' \|e\|.
\end{align}
This shows that the ISSf-BC holds where safety violation is minor (close to the original safe set~$\Cc$). Nevertheless, we cannot yet establish safety on $\Dc$ because there is no guarantee that $\Dc$ will be in the neighborhood of $\Cc$. For instance, by adjusting the parameter $d$ from zero, the set $\Dc$ may depart from $\Cc$ unboundedly to infinity. In such a case, however, it simply means the barrier function $h$ is not a good measure for minor safety violations and is unsuitable for formalizing practical safety. Thus, we make the following assumption.
\begin{assumption}\longthmtitle{Proximal Level Curves}\label{assumpion: proximal}
    The level curves of $h$ are within a certain distance from its zero level curve~$\partial \Cc$. That is, given a constant $d >0$ such that $h^{-1}(d)$ is nonempty, there exists an upper bound $\de>0$ such that  $\|x\|_{\partial \Cc}<\de$ for all $x\in h^{-1}(-d)$.~\hfill$\bullet$ 
\end{assumption}
The above assumption requires that $h$ is a suitable measure for minor safety violations, as well as practical safety. Indeed, the idea is related to ideas such as upper semi-continuity of level sets~\cite{AJ-F:Set:09, RK-AA-SG:19} and coercivity~\cite{AT-VD-RK-YY-AA:22}, which can be obtained under mild conditions. We omit those details and simply make the above assumption in the interest of space and to avoid drifting away from the main narrative with mathematical details.
The following theorem formalizes the notion of minor safety violations assumed in practical safety:
\begin{theorem}\longthmtitle{Forward Invariance of Expanded Set $\Dc$}\label{expanded set theorem}
    Consider the sample-and-hold control system~\eqref{eq:NL_sampled}. Under Assumptions \ref{Lipschitz controller}-\ref{assumpion: proximal}, for each $d>0$, there exists an upper bound~$e_{\max}>
    0$ on the sample-and-hold error such that, if $\|e\|_\infty \leq e_{\max}$, the set $\Dc$ given in~\eqref{eq:expanded set} is forward invariant.
\end{theorem}
\begin{proof} 
We begin by noting that under Assumption \ref{assumpion: proximal}, there exists a $\de>0$ such that $\lambda'=M\de+\lambda$ is the upper bound on $\Lie_gh$ on set $\Dc$, and thus, \eqref{eq:lambda prime}~holds. Defining $\emm = -\alpha(-d)/L_k\lambda'$, we have:
 \begin{align*}
\dot{h_d}(x,k_\nom&(x+e))=\dot{h}(x,k_\nom(x+e))\\
    &\geq -\alpha(h(x)) - L_k\lambda' \emm\\
    &= -\alpha(h_d(x)-d) + \alpha(-d)= -\alpha_d(h_d(x))
\end{align*}
where $\alpha_d(r) = \alpha(r-d)-\alpha(-d)$. Because $\alpha_d(0) = 0$ and $\alpha$ is a class-$\Kc_\infty^e$, $\alpha_d$ is also a class-$\Kc_\infty^e$. Thus, the expanded set $\Dc$ is forward invariant, concluding the proof.
\end{proof}
The theorem above justifies expanding the set $\Cc$ to accommodate small sample-and-hold errors. Our result is based on the ISSf concept, particularly through the ISSf-BC in~\eqref{eq:lambda prime}. We note importantly, however, that our ISSf-BC~\eqref{eq:lambda prime} only holds on $\Dc$ and cannot accommodate arbitrarily large $\|e\|$ even if we select $d$ to be large, unlike the standard ISSf. This limitation arises because $\lambda'$ depends on $d$, and as a result, our function $\iota$ in \eqref{eq:ISSf-CBF} is not uniform for all $d$. Fortunately, we avoid arbitrarily large sample-and-hold errors through high frequency sampling, which we discuss next.

\subsection{Small Error Bounds}
Our next step is to establish a relationship between the error $\|e\|$ and time between two sampling instants. Recall that at each instant $t_i$, the controller is sampled, and $e(t_i) = 0$ by definition. It follows intuitively that if only a short time elapses before the next sampling instant $t_{i+1}$, then $\|e\|$ should not grow much. This relies on the following assumption: 

\begin{assumption}
\longthmtitle{Bounded Dynamics}\label{bounded dynamics}
The dynamics describing sample-and-hold control-affine system~\eqref{eq:NL_sampled} are bounded on a given safe set $\Cc$, i.e. $\|\dot{x}\| = \|f(x) + g(x)k(x+e)\|\leq B_e$ for all $x\in\Cc$ with a positive constant $B_e$.~\hfill$\bullet$  
\end{assumption}
Once again, the assumption on boundedness is justified by the fact that it holds for compact safe sets (or compact regions of operation, cf. Remark \ref{rmk:regularity}).
From a similar argument as $\Lie_gh$, it follows that under Assumption \ref{Lipschitz controller}, \ref{assumpion: proximal} and \ref{bounded dynamics}, there exists a bound $B_e' \geq B_e$ on the set $\Dc$, providing a sufficient condition for establishing a sample-and-hold error bound.

\begin{lemma}\longthmtitle{Bounded Error}\label{bound on error Lemma-new} Consider the sample-and-hold system~\eqref{eq:NL_sampled}. Let $T_s=t_{i+1}-t_i$ denote the inter-event time between two sampling instants. Under Assumption \ref{Lipschitz controller}, \ref{assumpion: proximal}, and \ref{bounded dynamics}, the sample-and-hold error is bounded as $\|e(t)\| \leq B_e'T_s$ for all $t \in [t_i,t_{i+1})$ if the states along the trajectory $t \mapsto x(t)$ remain in the expanded set $\Dc$.

\end{lemma}
\begin{proof} From $e(t) = x(t_i) - x(t)$, we have $\dot{e}(t) = -\dot{x}(t)$. Then, by considering the evolution of $\|e(t)\|$, because the state remains in the set $\Dc$, the following bound holds:
     \begin{align*}
    \|e(t)\| &= \Big\|\overbrace{e(t_i)}^0 + \int_{t_i}^{t} \!\!\overbrace{-(f(x(\tau))+g(x(\tau))k_\nom(x(t_i)))}^{\dot{e}(\tau)} d\tau\Big\| \nonumber\\ 
             &\leq \int_{t_i}^{t}B_e'd\tau = B_e'(t-t_i).
    \end{align*}
    Given the inter-event time $T_s = t_{i+1}-t_i$, we obtain $\|e(t)\| \leq B_e'T_s$ for all time $t\in[t_i,t_{i+1})$, concluding the proof. 
\end{proof}
These constructions enable us to formalize practical safety.
\subsection{Formalizing Practical Safety}

We combine Theorem \ref{expanded set theorem} and Lemma \ref{bound on error Lemma-new} to formalize practical safety with the following result:
\begin{theorem}\longthmtitle{Practical Safety via ISSf}
    Consider the sample-and-hold control system~\eqref{eq:NL_sampled} with periodic sampling instants $\{t_i\}_{i=0}^\infty$, i.e., $t_{i+1}-t_i= T_s$ for all $i\in\naturals$. Under Assumptions \ref{Lipschitz controller}-\ref{bounded dynamics}, for each $d>0$, there exists a sampling time $T_s>0$ small enough such that the expanded set $\Dc$ given in~\eqref{eq:expanded set} is forward invariant.
\end{theorem}

\begin{proof}
    We begin by substituting the error bound from Lemma \ref{bound on error Lemma-new} into the ISSf-BC in \eqref{eq:lambda prime} to obtain:
     \begin{align*}
     \dot{h}_d(x,k_\nom(x+e)) &\geq -\alpha(h(x)) - L_k\lambda' \|e\| \nonumber \\
     &\geq -\alpha(h(x)) - L_k\lambda' B_e' T_s  \nonumber \\
     &= -\alpha(h_d(x) - d) - L_k\lambda' B_e' T_s  \nonumber \\
     &= -\alpha_d(h_d(x)) - \alpha(-d) - L_k\lambda' B_e' T_s
    \end{align*}
    where $\alpha_d$ is defined in Theorem \ref{expanded set theorem}. From the expression above, we can see that any sampling time satisfying
    $T_s \leq \frac{-\alpha(-d)}{L_k \lambda' B_e'}$
    ensures $\dot h_d(x,k_\nom(x+e)) \geq -\alpha_d(h_d(x))$, rendering $\Dc$ forward invariant. This concludes the proof.
\end{proof}
We have formalized the concept of practical safety within the framework of Input-to-State Safety (ISSf). By leveraging the key assumptions underlying practical safety, we have shown that the safety of sample-and-hold control systems can be established on an expanded set that accommodates the errors by allowing safety violations on the original safe set. However, we explore an alternative strategy in the following section that guarantees safety on the original safe set, avoiding any violations.
\section{Violation-Free Sample-and-Hold Safety}\label{violation free section}
This section presents an alternative strategy of achieving safety of the sample-and hold system \eqref{eq:NL_sampled} given a desired sampling time. Formalizing practical safety with the ISSf concept in the previous section provides tools we can leverage to make safety guarantees on the original safe set $\Cc$. While typically, ISSf involves expanding sets to accommodate errors (\ie violations of the original set), we explore an alternative approach of guaranteeing safety without resorting to the expansion of the set $\Cc$. In particular, we make use of the nominal controller~$k_\nom$ and leverage the fact that it satisfies the ISSf-BC~\eqref{eq:ISSf-lambda-final}. Our strategy introduces a robustness term (without expanding the safe set) in the ISSf-BC to deal with the sample-and-hold error.
Specifically, we propose the following adjustment of the nominal controller:
    \begin{align}\label{eq: K_T}
        k(x) = k_\nom(x) + \frac{1}{\vr}\Lie_gh(x)^\top,
    \end{align}
with a tuning parameter $\vr >0$ to be specified. The term $\frac{1}{\vr}\Lie_gh(x)^\top$ adjusts a nominal controller in the direction that increases the barrier function $h$, which will provide robustness against sample-and-hold errors. However, the newly added term itself will also suffer from errors due to sample-and-hold, and so careful analysis is needed to assess the full benefit of the adjustment.
\subsection{Safety from Controller Adjustment}
We begin by first acknowledging that the added term in the controller~\eqref{eq: K_T} is only useful when $\Lie_gh$ is nonzero. Fortunately, we only require this condition at the boundary of the safe set.

\begin{assumption}\longthmtitle{Control Authority on the Boundary of the Safe Set} \label{set boundary assumption}
The function $\Lie_gh$ is bounded from below on the boundary of the safe set $\partial\Cc$. That is, there exists a positive constant $\mu>0$ such that  $\|\Lie_gh(x)\| \geq \mu$ for all $x \in \partial \mathcal{C}$.~\hfill$\bullet$
\end{assumption}
This assumption is related to the possibility of rendering any safe set forward invariant in the presence of sample-and-hold errors. For states $x$ where $\Lie_gh(x)=0$, the barrier function~$h$ evolves at the mercy of the drift $f$, as there is no way of influencing it with any control input. Thus, there is no guarantee that the set will remain forward invariant under sample-and-hold control. In essence, this assumption ensures the well-posedness of the problem we set out to solve.
Given the above assumptions and the adjusted controller, we establish the existence of a sampling time that guarantees the safety of \eqref{eq:NL_sampled} on $\Cc$ in the following result.
%
%
\begin{theorem}\label{main theorem-new} \longthmtitle{Safety on Original Safe Set with Adjusted Controller}
Consider the sample-and-hold control system \eqref{eq:NL_sampled} using the adjusted controller provided in \eqref{eq: K_T} instead of $k_\nom$. Let the controller be sampled periodically with a sampling time $T_s$, then under the Assumptions \ref{Lipschitz controller}-\ref{Lipschitz assumption Lgh-new}, \ref{bounded dynamics} and \ref{set boundary assumption}, there exists a small (fast) enough sampling time $T_s$ such that the following condition holds:
\begin{align}\label{eq:main theorem hdot 1}
    \dot{h}(x,k(x+e)) \geq &-\alpha(h(x)),
\end{align}
for states $x$ on a small $\de$-neighborhood of $\partial \Cc$. Hence, the safe set
$\Cc$ is forward invariant.
\end{theorem}
\begin{proof}
We begin by noting that under Assumptions \ref{Lipschitz assumption Lgh-new} and \ref{set boundary assumption}, there exists a nonzero lower bound $\mu' \leq \mu$ for the function $\|\Lie_gh\|$, within the $\delta$-neighborhood of the set boundary $\partial \Cc$. We can establish the following ISSf-BC on $\partial \Cc$:
\begin{align} \label{eq: h_dot sample and hold adjusted-new}
\dot{h}&(x,k(x+e)) \nonumber\\
&= \Lie_fh(x)+\Lie_gh(x)(k_\nom(x_i)+\frac{1}{\vr}\Lie_gh(x_i)^\top) \nonumber \\
                &\geq -\alpha(h(x)) - \|\Lie_gh(x)\| L_k\|e\| +\frac{1}{\vr}\Lie_gh(x)\Lie_gh(x_i)^\top \nonumber \\
                &=-\alpha(h(x)) - \|\Lie_gh(x)\| L_k\|e\|+\frac{1}{\vr}\|\Lie_gh(x)\|^2\nonumber \\
                &\quad  +\frac{1}{\vr}\Lie_gh(x)(\Lie_gh(x_i)-\Lie_gh(x))^\top\nonumber \\
                %
                %
                &\geq -\alpha(h(x)) + \|\Lie_gh(x)\|\Big( \frac{1}{\vr}\|\Lie_gh(x)\| -(L_k+\frac{M}{\varepsilon} )\|e\|\Big) \nonumber\\
                &\geq -\alpha(h(x)) + \|\Lie_gh(x)\|\Big( \frac{\mu'}{\vr} -(L_k+\frac{M}{\varepsilon} )\|e\|\Big), 
\end{align}
where we have used the shorthand notation $x_i = x(t_i)$. Next, similar to the result of Lemma \ref{bound on error Lemma-new}, under Assumption \ref{bounded dynamics}, we can establish an error bound $\|e(t)\| \leq BT_s$ where $B$ is the bound on our dynamics \eqref{eq:NL_sampled} with the controller \eqref{eq: K_T}, instead of $k_\nom$. 
This leads to the following:
\begin{align} 
\dot{h}(x,&k(x+e)) \\
&\geq -\alpha(h(x)) + \|\Lie_gh(x)\|\Big( \frac{\mu'}{\vr} -(L_k+\frac{M}{\varepsilon} )BT_s\Big). \nonumber
\end{align} 
By picking $T_s < \mu'/(B(\vr L_k+M))$ to make the last term positive, the barrier condition
$\dot{h}(x,k(x+e)) \geq -\alpha(h(x))$ is met on $\partial \Cc$, i.e., $\dot h (x,k(x+e)) \geq 0$ where $h(x)=0$. In addition, due to Assumption~\ref{set boundary assumption}, $\frac{\partial h}{\partial x}\neq 0$ on $\partial \Cc$, so by Nagumo's theorem~\cite{FB:99}, the set $\Cc$ is forward invariant.
\end{proof}
The above theorem establishes a violation-free inter-sampling result. The result shows that with an adjustment to the controller, the sample-and-hold system can be rendered safe on the original safe set~$\Cc$, with a high enough sampling frequency. However, this result does not provide any barrier condition~\eqref{eq:CBF} for every state $x$ in the safe set $\Cc$. This may be problematic as many safety-related concepts in the literature may rely on such a condition. One particular example is event-triggered control, which we discuss in Section \ref{sec: Event trigger}.

\subsection{Barrier Condition for Adjusted Controller}
In order to derive a barrier condition for the entirety of the safe set $\Cc$, we note one key obstruction. That is, there is a possibility that $\Lie_gh$ may approach zero for states at least $\de$ away from the boundary of the safe set, i.e., for $x\in\Cc$ such that $\|x\|_{\partial \Cc}>\de$. In such a case, our adjustment in \eqref{eq: K_T} to the nominal controller is ineffective. Nevertheless, we can use the following fact:
\begin{align}\label{alpha prime}
\alpha(h(x)) &= (1+c)\alpha(h(x))-c\alpha(h(x)) \nonumber \\
&= \alpha'(h(x))-c\alpha(h(x))
\end{align}
for any positive constant $c>0$, and the function $\alpha'= (1+c)\alpha$ is also a class-$\Kc_\infty^e$ function. The idea is to leverage the fact that $h(x)\neq 0$ inside the safe set, to use $c\alpha(h(x))$ to provide the necessary robustness against the sample-and-hold error. To this end, we make the following assumption

\begin{assumption}\longthmtitle{Lower Bound on $h$}\label{small h assumption}
For any given $\de>0$, there exists $h_{\m}>0$ such that $h(x)\geq h_{\m}$ 
for all $x\in \Cc$ such that $\|x\|_{\partial \Cc}\geq\delta$.~\hfill$\bullet$ 
\end{assumption}
The assumption is based on the idea that barrier function $h$ should be a proper safety measure, ensuring it has higher values farther away from unsafe states. 
The property \eqref{alpha prime} and the controller~\eqref{eq: K_T} ultimately lead to a barrier condition for every $x \in \Cc$, which we present in the following theorem:

\begin{theorem}\label{main theorem 2} \longthmtitle{Barrier Condition on Original Safe Set with Adjusted Controller}
Consider the sample-and-hold control system \eqref{eq:NL_sampled} using the adjusted controller provided in \eqref{eq: K_T} instead of $k_\nom$. Let the controller be sampled periodically with a sampling time $T_s$, under the Assumptions \ref{Lipschitz controller}-\ref{Lipschitz assumption Lgh-new} and \ref{bounded dynamics}-\ref{small h assumption}, there exists a small (fast) enough sampling time $T_s$ such that the following condition holds:
\begin{align}\label{eq:main theorem hdot 2}
    \dot{h}(x,k(x+e)) \geq &-\alpha'(h(x)),
\end{align}
for all $x \in \Cc$. Hence, the safe set
$\Cc$ is forward invariant.
\end{theorem}

\begin{proof}
From Theorem~\ref{main theorem-new}, we note that:
$$
\dot h(x,k(x+e)) \geq -\alpha(h(x))\geq -\alpha'(h(x)),
$$
on a small $\de$-neighborhood under the assumption of the theorem. Therefore, we will focus on states $x$ such that $\|x\|_{\partial \Cc}\geq\de$. To this end, Assumption~\ref{small h assumption} dictates that $h(x)\geq h_{\m}$, so
we have $\alpha(h(x)) \geq \alpha(h_{\m})$.
Following the proof of Theorem~\ref{main theorem-new}, we obtain the following inequality:
\begin{align*}
\dot{h}(x,k(x+e&)) \geq -\alpha'(h(x)) + c\alpha(h_{\m})+  \frac{\|\Lie_gh(x)\|^2}{\vr} \nonumber \\ & \ \quad \quad  -\|\Lie_gh(x)\|\Big((L_k+\frac{M}{\varepsilon} )BT_s\Big) \nonumber\\
&\geq -\alpha'(h(x)) + c\alpha(h_{\m})-\lambda\Big(L_k+\frac{M}{\varepsilon} \Big)BT_s.\nonumber
\end{align*} 
Thus, we establish a safe sampling time by picking a sampling time satisfying:
\begin{align}\label{eq:MIET}
    T_s < \min \Bigg\{\frac{c\alpha(h_{\m})}{\lambda(L_k + \frac{M}{\vr})}  \ , \ \frac{\mu'}{B(\vr L_k+M)} \Bigg\},
\end{align}
we ensure that the barrier condition~\eqref{eq:main theorem hdot 2} is met for all $x \in \Cc$, concluding the proof.
\end{proof}
The above theorem establishes a barrier condition for all $x \in \Cc$ by ensuring that, with the adjusted controller~\eqref{eq: K_T}, robustness against sample-and-hold errors is always guaranteed for the system \eqref{eq:NL_sampled}. In contrast to the strategy of set expansions discussed in Section \ref{Set Expansion Section}, Theorem \ref{main theorem 2} ensures safety on the original set $\Cc$, avoiding violations. 

The main drawback of Theorem~\ref{main theorem 2} is its conservatism in the sampling time it guarantees. In presenting our theorem statement, we elect to omit the bound on the safe sampling time $T_s$ because it involves various constants $B, \lambda, \mu', M, L_k$, which are difficult to find in practice. In addition, our derivations are overly conservative at some key steps. For instance, the bound on sample-and-hold error in Lemma~\ref{bound on error Lemma-new} supposes the state deviates as fast as possible from the state where it is last sampled. To resolve this, we leverage the results in this section to propose an event-triggered approach that efficiently samples the controller.

\subsection{Violation-Free Event Triggered Safety}\label{sec: Event trigger}
Event-triggered control, as proposed in~\cite{PT:07},~\cite{AJT-PO-JC-AA:21-csl} offers a way to mitigate conservatism in controller sampling. In this framework, the control is sampled at time instants (\ie events) prescribed by a state-based criterion, rather than periodically based on time. 
Most importantly, in our context, we can leverage the barrier condition that we developed in Theorem \ref{main theorem 2} to establish a trigger condition that ensures the sample-and-hold system~\eqref{eq:NL_sampled} is safe $\Cc$. To this end, we use the following reasonable trigger condition to maintain safety:
\begin{align}\label{trigger T}
    \mathcal{T}(x,e) =  \dot{h}(x,k(x+e)) + \alpha'(h(x)).
\end{align}
 The event-triggered idea is to monitor \eqref{trigger T} to make sure it always remains positive, that is, $\mathcal{T}(x(t),e(t)) > 0$ along the trajectory and only to trigger a control sampling when it gets violated as:
\begin{align} \label{eq: ETC trigger}
    t_{i+1} = \min \big \{t \geq t_i: \mathcal{T}(x(t),e(t)) = 0 \big \}.
\end{align}
The main concern with event-triggered control is related to its aperiodic nature of controller samplings. More specifically, it becomes possible that an infinite number of controller samplings are triggered within a finite time period (Zeno behavior). 
Fortunately, the existence of a sampling time provided by Theorem \ref{main theorem 2} eliminates this possibility. We now formalize our event-triggered control strategy.
 \begin{theorem} \longthmtitle{Trigger Law for Safety on $\Cc$} \label{ETC Theorem}
Consider the sample-and-hold system~\eqref{eq:NL_sampled} with a adjusted controller given by~\eqref{eq: K_T}. Let the sequence of sampling time $\{t_i\}_{i\in \naturals}$ be determined iteratively according to the trigger law~\eqref{eq: ETC trigger}. Under Assumptions \ref{Lipschitz controller}-\ref{Lipschitz assumption Lgh-new} and \ref{bounded dynamics}-\ref{small h assumption}, there exists a minimum inter-event time $T_s >0$ such that~$t_{i+1}-t_i \geq T_s$ for all $i\in \naturals$. Consequently, the barrier condition~\eqref{eq:main theorem hdot 2} is satisfied for all time along the trajectory, and the set $\Cc$ is rendered forward invariant.  
 \end{theorem}

\begin{proof}
Under the assumptions, we can follow the proof of Theorem \ref{main theorem 2} to guarantee the existence a $T_s>0$ that ensures the quantity~$\mathcal{T}(x(t),e(t))$ cannot become negative before $T_s$ has elapsed. It can then be concluded that $t_{i+1} - t_i \geq T_s$ for all $i\in\naturals$. Combining this with the fact that the trigger condition~\eqref{eq: ETC trigger} ensures that $\mathcal{T}(x(t),e(t)) > 0$ ensures that $\dot{h}(x(t),k(x(t)+e(t))) > - \alpha'(h(x(t)))$ along the trajectories, and $\Cc$ is forward invariant.
\end{proof}

The theorem above establishes an event-triggered strategy that ensures the safety of the sample-and-hold control system \eqref{eq:NL_sampled} on the set $\Cc$. Moreover, there is no need to specify the various bounds to obtain a suitable sampling time as required in Theorem \ref{main theorem 2}. Instead, to utilize event-triggered control, we note that any positive values of $c > 0$ and $\vr >0$ are sufficient, albeit small values of these parameters will result in a higher sampling rate while large values will result in lack of robustness or larger control input, respectively.
We now illustrate the effectiveness of our violation-free safe sampling design through simulation experiments.

\section{Application To Adaptive Cruise Control} \label{Application}
We apply our results to the adaptive cruise control system, where a host vehicle is controlled to follow a lead vehicle while maintaining a safe distance with it's leader. We define $x = (x_1,x_2, x_3)$, where $x_1$ represents the position of the host vehicle, $x_2$ represents its velocity, and $x_3$ represents the headway distance between the host and the lead vehicle. We adopt the following point-wise model for the dynamics \cite{ADA-JWG-PT:14}:
\begin{eqnarray*}
\dot x  
=  
\begin{bmatrix}
x_2   \\ 
-\frac{1}{m}F_r \\
v_0 - x_2
\end{bmatrix}+ 
\begin{bmatrix}
 0 \\ \frac{1}{m} \\ 0 \\ 
 \end{bmatrix}u \, ,
\end{eqnarray*}
where $F_r = f_0 + f_1 x_2 + f_2 x_2^2 $ represents the resistance force (in Newtons $N$) experienced by the host vehicle while $m$ is it's mass. $v_0$ denotes a constant velocity of the leading vehicle. Motivated by \cite{ADA-JWG-PT:14}, we take $m = 1650 \, kg, v_0 = 13.89 \, \frac{m}{s}, f_0=0.1 N, f_1=5 \frac{Ns}{m}$ and $f_2=0.25 \frac{Ns^2}{m}$.
\subsection{Nominal Controller}
We use a nominal controller to direct the host vehicle towards a desired objective, which is a specified speed ($v_d = 22 \frac{m}{s}$ in our case). When the host vehicle enters unsafe regions, a safety controller will take over by minimally adjusting the nominal control input. To implement the desired objective, we employ the desired controller detailed in \cite{ADA-JWG-PT:14}:
 \begin{align}
    u_{des}(x) =  -\bar{\varepsilon} \frac{m}{2}(x_2 -v_d) + F_r(x),
 \end{align}
  where $\bar{\varepsilon} = 10$ is presented in \cite{ADA-JWG-PT:14} as the decay rate necessary for the stability of $x_2$ to $v_d$.
 \subsection{Safety Controller}
In contrast to the approach taken in \cite{ADA-JWG-PT:14}, we adopt a nonlinear model structure for time headway by incorporating a variable time headway \cite{CW-ZW-YL-CF-KL-MH:20} in the barrier function:
$$ h(x) = x_3 - T_h(x) x_2.$$ 
Here, $T_h(x) = 1.8x_2$ varies depending on the host vehicle's speed and a constant value of 1.8. By taking the derivative of our barrier function, we obtain:
$$ \dot{h}(x,u) = (v_0-x_2) + \frac{3.6x_2F_r}{m} - \frac{3.6x_2}{m}u$$ 
To ensure safety, we implement a Quadratic Program (QP) based safety controller that minimally adjusts the desired control input $u_{des}$ through a safety filter \cite{ADA-SC-ME-GN-KS-PT:19}:
\begin{align*}\label{eq:CLF-QP}
k_{\nom}(x) &= \argmin_{u \in \mathbb{R}}||u - u_{des}(x)||^2 \tag{CBF-QP}\\
&\text{s.t.}\ \dot{h}(x,u) \geq -\alpha(h(x)).
\end{align*}

\begin{figure}[t!]
\centering
\includegraphics[width=1\linewidth,height=\textheight,keepaspectratio]{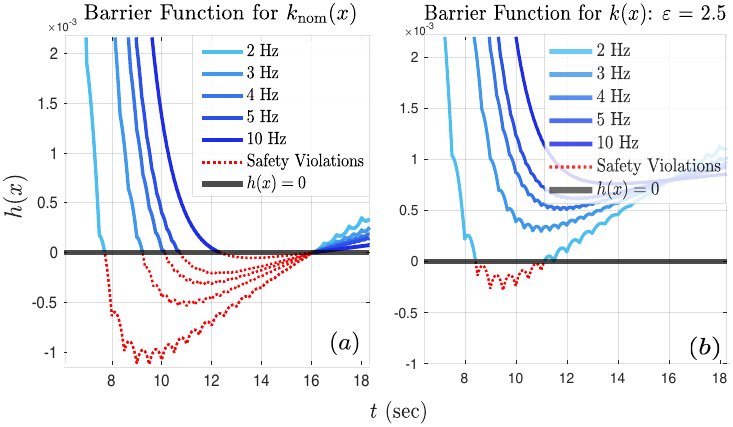}
\caption{Illustration of the relationship between sampling frequency and $\varepsilon$ and their satisfaction of safety, initialized from a positive value of $h(x)$.}
\label{final figure}
\vspace{-5mm}
\end{figure}
\begin{figure*}[t!]
\label{ETC figs}
\centering
{\includegraphics[width=0.34\linewidth]{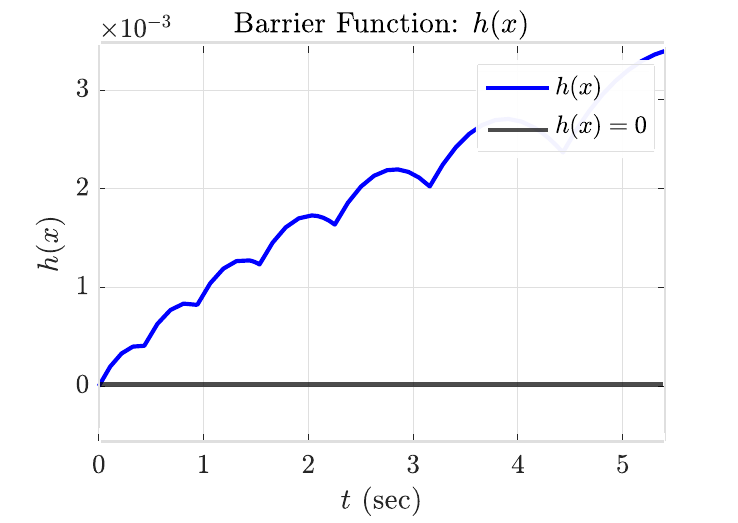}\hspace{-10pt}}%
{\includegraphics[width=0.3\linewidth]{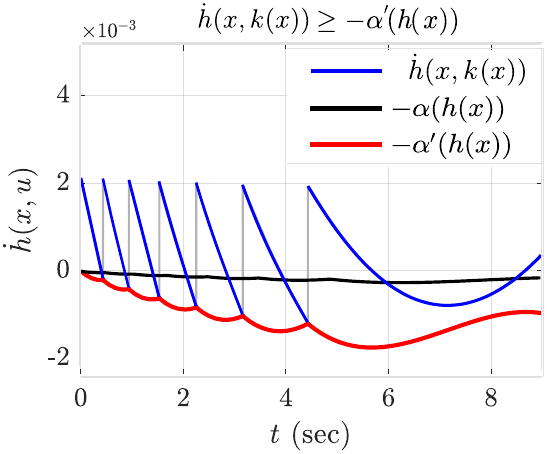}\hspace{2pt}}%
{\includegraphics[width=0.31\linewidth]{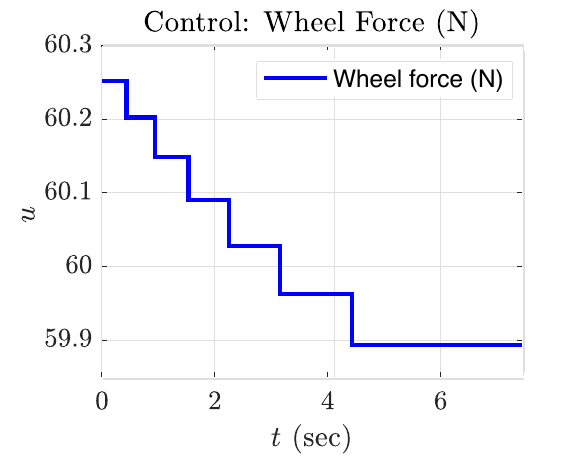}}%
\caption{Event Triggered Control results showing (from left to right): barrier function values, trigger condition satisfaction, and sampled control values.}
\vspace{-4mm}
\end{figure*}
\vspace{-10pt}
\subsection{Simulation Results}
We adjusted the the nominal controller $k_{\nom}(x)$ from the CBF-QP, as presented in \eqref{eq: K_T} and observe it's capability of achieving safety. In Figure 2, we present the results of our implementation using various sampling frequencies, with $c = 9.18$, $h_{\m} = 0.0005$, and $\varepsilon = 2.5$. It can be seen that slower sampling frequencies result in safety violations. However, by utilizing Theorems \ref{main theorem-new} and \ref{main theorem 2}, we are able to achieve safety for the original safe set $\Cc$ at fast enough sampling frequencies. Additionally, compared to Figure 1., we leverage Theorem \ref{ETC Theorem} to demonstrate how the event-triggered control strategy achieves less conservative sampling while still ensuring the safety of our original safe set in Figure 3.

\section{Conclusion}
We have formalized practical safety through the Input-to State Safety framework. From this new perspective, we have proposed an alternative method of achieving sample-and-hold safety without any set violations (i.e., set expansions). Our approach involves adjusting the nominal controller to provide robustness near the boundary of the safe set. Furthermore, our analysis has enabled us to leverage event-triggered control for safety, and thus, we can mitigate the conservatism in choosing a sampling frequency.  Future work involves the extension of our work to study the effects of measurement errors associated with the sampling of safeguarding controllers. 

\vspace{0.2cm}
\noindent \textbf{Acknowledgement.} The authors thank Ryan K. Cosner and Andrew J. Taylor for their discussions and insightful perspectives on sampled data theory and practical safety.



\bibliography{alias,PO,GB, main-GB, cosner}
\bibliographystyle{ieeetr}

\end{document}